\documentclass[twocolumn]{aastex63} 
\usepackage{amsmath}
\usepackage{mathtools}
\usepackage{lineno}

\newcommand{\mearth}{M$_\oplus$}

\shorttitle{Spin Dynamics of `Oumuamua}
\shortauthors{Seligman et al.}

\graphicspath{{./}{figures/}}

\begin{document}

\title{On The Spin Dynamics of Elongated Minor Bodies with Applications to a Possible Solar System Analogue Composition for `Oumuamua}

\correspondingauthor{Darryl Seligman}
\email{dzseligman@uchicago.edu}

\author[0000-0002-0726-6480]{Darryl Seligman}
\affiliation{Dept. of the Geophysical Sciences, University of Chicago, Chicago, IL 60637}

\author[0000-0002-1422-4430]{W. Garrett Levine}
\affil{Dept. of Astronomy, Yale University, 52 Hillhouse, New Haven, CT 06511, USA}
\author[0000-0001-9749-6150]{Samuel H. C. Cabot}
\affil{Dept. of Astronomy, Yale University, 52 Hillhouse, New Haven, CT 06511, USA}
\author[0000-0002-3253-2621]{Gregory Laughlin}
\affil{Dept. of Astronomy, Yale University, 52 Hillhouse, New Haven, CT 06511, USA}

\author[0000-0002-2058-5670]{Karen Meech}
\affil{Institute for Astronomy, University of Hawaii, 2680 Woodlawn Drive, Honolulu, HI 96822, USA}

\begin{abstract}

The first interstellar object, 1I/2017 U1 (`Oumuamua), exhibited several unique properties, including an extreme aspect ratio, a lack of typical cometary volatiles, and a deviation from a Keplerian trajectory. Several authors have hypothesized that the non-gravitational acceleration was caused by either cometary outgassing or radiation pressure. Here, we investigate the spin dynamics of `Oumuamua under the action of high surface area fractional activity and radiation pressure. We demonstrate that a series of transient jets that migrate across the illuminated surface will not produce a secular increase in the spin rate. We produce 3D tumbling simulations that approximate the dynamics of a surface covering jet, and show that the resulting synthetic light curve and periodogram are reasonably consistent with the observations. Moreover, we demonstrate that radiation pressure also produces a steady spin-state. While carbon monoxide (CO) has been dismissed as a possible accelerant because of its non-detection in emission by \textit{Spitzer}, we show that outgassing from a surface characterized by a modest covering fraction of CO ice can satisfy the non-ballistic dynamics for a plausible range of assumed bulk densities and surface albedos. \textit{Spitzer} upper limits on CO emission are, however, inconsistent with the CO production necessary to provide the acceleration. Nonetheless, an ad hoc but physically plausible explanation is that the activity level varied greatly during the time that the trajectory was monitored. We reproduce the astrometric analysis presented in \citet{Micheli2018}, and verify that the non-gravitational acceleration was consistent with stochastic changes in outgassing.

\end{abstract}


\section{Introduction}

1I/2017 U1 (`Oumuamua), which achieved notoriety as the first macroscopic object  discovered to traverse the Solar System from interstellar space, bore scant resemblance to any known astronomical object. Following its arrival from the galactic apex \citep{mamajek2017} and approach within 0.25 au of the Sun, its brightness was seen to vary quasi-periodically by more than a factor of ten, and with a duty cycle of roughly 3.6 hours \citep{Meech2017}, implying an extreme aspect ratio \citep{Jewitt2017,bannister2017col, Knight2017,fra18, dra18}. Extensive modeling of the light curve has found that a tumbling oblate ellipsoid of aspect ratio 6:6:1 \citep{mashchenko2019modelling} provides a good fit to the photometric data. During the observational campaign, `Oumuamua was found to lack a detectable dust coma, suggesting a refractory composition despite its red color matching that of volatile-rich comets \citep{Meech2017,Jewitt2017}. Later, however, upon analysis of astrometric data, \cite{Micheli2018} determined that `Oumuamua's motion deviated from a purely ballistic trajectory, finding that the component of non-Keplerian acceleration was consistent with $4.9\times10^{-4} (r/1\rm{au})^{-2}\,\bf{\hat{r}}\, \rm{cm\,s^{-2}}$, a value roughly 1/1000$^{\rm {th}}$ the magnitude of the solar gravitational acceleration, where $\bf{r}$ denotes the heliocentric distance. Given the lack of measured outgassing and the absence of a detectable tail of micron-sized dust \citep{Micheli2018}, the driver of the non-Keplerian trajectory, as well as the object's composition and albedo, remain unknown.

`Oumuamua's exceptional physical attributes have led to abundant and varied proposals that place it as the first-detected member of a larger population. Generally, these interpretations satisfy the non-detection of a coma and non-ballistic trajectory with propulsion by either solar radiation pressure or outgassing of atypical sublimated ices. The former mechanism was initially  suggested by \citet{Micheli2018}, and  \cite{bialy2018could} explored it in the context of a millimeter-thick, lightsail-like object. Models such as those presented by \cite{MoroMartin2019} and \citet{luu2020oumuamua} consider the same driving force on ultra-porous dust aggregates.

Comets in the solar system frequently undergo non-gravitational acceleration driven by the sublimation of outer solar system ices such as H$_2$O, CO, and CO$_2$ \citep{Reach2013,Ahearn:1995,Micheli2018}. Therefore, a seemingly less exotic explanation for the non-gravitational acceleration would be cometary outgassing. \citet{Sekanina2019}, however, demonstrated the infeasibility of jetting H$_2$O for the non-gravitational acceleration due to water ice's high enthalpy of sublimation. Moreover, \cite{Seligman2020} confirmed that H$_2$O could not be the dominant accelerant and ruled-out CO$_2$.  H$_2$ ice was discussed as a possible constituent of `Oumuamua by \cite{fuglistaler2018solid} and reconciled with the energy balance of the non-gravitational acceleration by \cite{Seligman2020}. Molecular nitrogen ice was demonstrated to be consistent with the non-Keplerian trajectory by \cite{jackson20211i}.

If `Oumuamua's non-gravitational acceleration was powered by cometary outgassing or radiation pressure, then the resulting spin dynamics of the object must be consistent with the observations. \citet{Rafikov2018} pointed out that cometary outgassing should lead to a significant change in the angular momentum of `Oumuamua during the period over which it was monitored, although the periodicity in the light curve was constant. \citet{Seligman2019} demonstrated that cometary outgassing could explain the non-gravitational acceleration while maintaining the near-constant spin period if the activity was confined to near the sub-stellar point of the illuminated surface. Based on the energetics of the non-gravitational acceleration, \citet{Seligman2020} demonstrated that a moderate or high fraction of the surface of `Oumuamua would be required to be covered in volatile material for outgassing to explain the acceleration. Therefore, in this study we extend the analysis in \citet{Seligman2019} to investigate the spin dynamics of an elongated body with jets that cover the entire illuminated surface. In \S \ref{sec:spin}, we present analytic and numerical models of `Oumuamua's spin state resulting from a jet covering the entire illuminated surface. This regime is more realistic than the discrete continuously subsolar jet considered by \cite{Seligman2019} for volatiles that require high surface coverage fractions. We establish qualitatively similar behavior between these two limiting cases. In neither scenario does `Oumuamua experience secular spin-up, so high surface covering activity from an elongated body is fully consistent with the observed light curve. We adjust our model to recover the spin dynamics of a body under the action of radiation pressure, and show that the resulting spin state is also stable.

All of the previously-proposed, exotic interpretations for `Oumuamua's origin imply the widespread production of previously-undetected and unanticipated objects. Moreover, each model draws on novel astrophysical processes or environments. As a consequence, an explanation of `Oumuamua that invokes a bulk composition similar to  minor bodies in the solar system would be a welcome development. 
 As an application of our dynamical model, we examine the consistency of carbon monoxide (CO) outgassing with the observational data on `Oumuamua. Previously, this common astrophysical volatile had been excluded from consideration for `Oumuamua based on the reported non-detection by the \textit{Spitzer Space Telescope} \citep{Trilling2018}, however, in this paper there was a factor of $\sim116$ error in the production rate limit, which should have been higher. Given its presence in Solar System objects, and that it is one of the three most abundant volatiles in comets that drives comet activity, CO bears analyzing from an energetic perspective. Because the aforementioned data were obtained within a single 33-hour span, it does not directly rule-out a scenario in which `Oumuamua carried typical cometary ices dominated by CO and was subject to highly variable outgassing. The only possible way that CO could be the driver of `Oumuamua's acceleration is that the interstellar object's outgassing level was below the limit during the attempted \textit{Spizter} observation, but that its average sublimation rate satisfied the non-gravitational acceleration over the entire trajectory. 

We investigate the energy balance of CO outgassing in \S \ref{sec:energetics}, finding that this ice could reconcile `Oumuamua's magnitude of non-gravitational acceleration for a range of albedos and bulk densities. Nevertheless, there are still phenomena not easily explained by the CO hypothesis. In \S \ref{sec:challenges}, we assess the upper bound on CO production. While we revise-up the $3\sigma$ production constraint from \cite{Trilling2018}, we still find that `Oumuamua's CO activity would need to vary by a factor of 100 around the average value to simultaneously satisfy the \textit{Spitzer} non-detection and the non-Keplerian trajectory. In addition, `Oumuamua's inbound kinematics and its young inferred age are not readily reconciled with CO, and the complete absence of any detectable micron-sized dust would be curious.

Further corroborating the general workability of CO, however, \S \ref{sec:supportive_evidence_astrometry} demonstrates that the astrometric data cannot immediately differentiate between sporadic and continuous outgassing. In addition, we confirm the result from \cite{Micheli2018} that `Oumuamua did accelerate non-gravitationally. Next, in \S \ref{sec:supportive_evidence_galactic_reservoir} we quantify the possible size of the galactic mass reservoir for interstellar objects and show that CO ice bodies could satisfy the inferred number density of interstellar objects. In \S \ref{sec:solar_system_CO}, we discuss observations of CO in Solar System objects.

Finally, we discuss the implications that arise if `Oumuamua was enriched in CO in \S \ref{sec:discussion}, and we outline the expectations for the Vera Rubin Observatory if this is the case. While the \textit{Spitzer} data places severe upper bounds on `Oumuamua's CO production when it was at a $2\,\text{au}$ distance, we conclude that the energetic, dynamical, and astrometric evidence, along with the composition of Borisov, may make CO a compelling hypothesis to reconcile `Oumuamua with the current galactic census of minor bodies.

\section{Spin Dynamics of `Oumuamua-Like Interstellar Objects} \label{sec:spin}

\citet{Rafikov2018} demonstrated that cometary outgassing could lead to observable spin up of the body during the period over which it was monitored. The study argued that reactive torques similar to those seen on Solar System comets, averaged over the trajectory, would increase the angular momentum. \citet{Seligman2019} demonstrated that comet-like outgassing can explain the ballistic trajectory while maintaining the near-constant spin period that was observed. This analysis relies on the assumption that the outgassing jet was able to quickly mobilize along the surface of the body to track the substellar point of maximal solar illumination. In this construction, the equations of motion reduce to a one degree-of-freedom Hamiltonian, and the oscillations  of `Oumuamua are reminiscent of those of a pendulum.

If `Oumuamua's non-gravitational acceleration was caused by outgassing, it is possible that a moderate or high fraction of the surface was covered in volatile material. \citet{Seligman2020} demonstrated that H$_2$ or N$_2$ would be energetically-consistent as `Oumuamua's accelerant for a range of albedo and bulk density with an ice surface coverage of $5-100\%$. Therefore, we extend the point jet model of \citet{Seligman2019} to account for outgassing from the entire surface, which would be more realistic  if `Oumuamua was composed of N$_2$, as explored by \citet{jackson20211i} and \citet{desch20211i}. We demonstrate that this extended model produces a very similar spin state as the single point model. We infer that if `Oumuamua's outflow morphology was in a regime between these two extremes, its spin dynamics would also be consistent with the light curve's static observed periodicity.

\subsection{Two-Dimensional Restricted Problem}
\begin{figure*}[]
\begin{center}
\includegraphics[scale=0.6,angle=0]{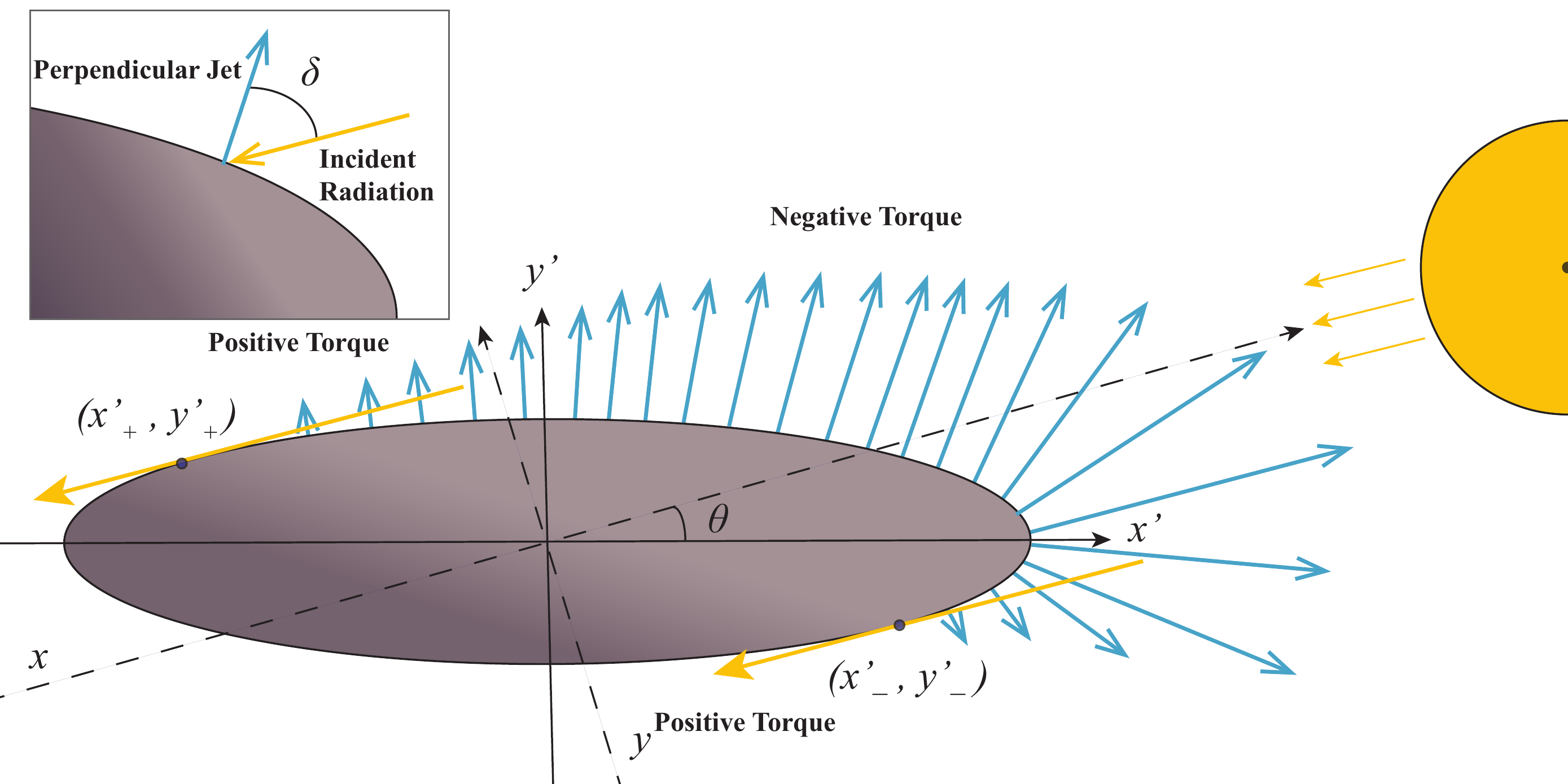}
\caption{Schematic diagram of outflows venting across the entire illuminated surface. The blue arrows show the position of a variety of jets, which act normal to the surface and whose strength is modulated by the angle $\delta$, which represents the angle between the normal and the direction to the incident solar radiation. The limits of illumination  are indicated with purple points, where the incident solar radiation is tangent to the slope of the ellipsoid. For any orientation of `Oumuamua, the torques in quadrants 2 and 4 will counteract the dominant torques in quadrant 1. \label{fig:schematic}}
\end{center}
\end{figure*}

In this section, we derive the equations of motion in an idealized 2D geometry, in which they reduce to a one degree of freedom Hamiltonian. We work in a non-inertial frame that co-moves with `Oumuamua, as illustrated in Figure \ref{fig:schematic}. The origin denotes the center of mass of `Oumuamua, and this frame is denoted with unprimed coordinates. The vector $\hat x$ is the direction of the radial vector connecting the Sun to the primed origin. We  consider the restricted geometry in which one axis of the ellipsoid is aligned with $\hat z$ and  the surface covering jet only acts in the $x$-$y$ plane. The angle $\theta\in(0,2\pi)$ represents a rotation about the $z$-axis, and defines a rotation matrix that represents a transformation from the  solar oriented unprimed frame to the body frame of the ellipse. The $\hat{x'}$ axis is along the major axis of `Oumuamua.  The equation for the ellipse in the body frame is,

\begin{equation}
    f(x',y')=\frac{x'^2}{a^2}+\frac{y'^2}{b^2}-1=0 \, ,
\end{equation}
where $a$ and $b$ are the semi-major and minor axes of the ellipse. In order to calculate the torque for a given rotation angle $\theta$, from a series of jets acting on the entirety of the illuminated surface of the ellipsoid, we must integrate along the surface of the ellipse. We therefore introduce the variable $\psi$, a non-physical angle that conveniently parameterizes the ellipse as

\begin{equation}
       x' = a\cos{\psi} \\ \,, \,
     y' = b\sin{\psi}\\\,,\,
   \psi\in(0,2\pi)\,.
\end{equation}
The limits of the contour integral correspond to the points on the surface of the ellipse that bound the solar illumination. These points are defined where the vector tangent to the surface of the ellipse is parallel to the vector to the sun. This occurs where
\begin{equation}
    \frac{\partial y}{\partial x}=\frac{\partial y}{\partial x'}\frac{\partial x'}{\partial x}+\frac{\partial y}{\partial y'}\frac{\partial y'}{\partial x}=0\,. 
\end{equation}
Solving this implicit equation for solution sets of the form $(x',y')$ yields a fourth-order polynomial with eight solutions, four of which are imaginary. Defining the variables $x'_\pm$ and $y'_\pm$ as
\begin{equation}
   x'_\pm=\pm\frac{a\sin\theta }{\sqrt{\epsilon^2\cos^2\theta+\sin^2\theta}}\\\,\,,\,
    y'_\pm = \pm \frac{b\epsilon\cos\theta}{\sqrt{\epsilon^2\cos^2\theta+\sin^2\theta}}\, ,
\end{equation}
the real solutions are $(x'_+,y'_+)$, $(x'_-,y'_-)$, $(x'_+,y'_-)$ and $(x'_-,y'_+)$, where $\epsilon=b/a$ is the aspect ratio. Mathematically these solutions are all extrema, but two of the solutions are antipodes of the physical limits. The two limits are the solutions  $(x'_+,y'_+)$ and $(x'_-,y'_-)$.

The unit normal to this ellipsoid in the body frame is
\begin{equation}
\begin{split}
    \frac{{\bf{\nabla}}f}{|{\bf{\nabla}}f|}=
    \bigg(\frac{x'^2}{a^4}+\frac{y'^2}{b^4}\bigg)^{-1/2}
    \bigg(\frac{x'}{a^2}\hat{\bf x'}+\frac{y'}{b^2}\hat{\bf y'}\bigg) \\
    =\frac{\epsilon\cos{\psi}\,\hat{\bf x'}+\sin{\psi}\,\hat{\bf y'}}{\sqrt{\sin^2{\psi}+\epsilon^2\cos^2{\psi}}}
    \, .\label{gradientEq2d}
\end{split}
\end{equation}
We introduce the zenith angle, $\delta(\psi)$, as the angle between the normal to the surface and the direction to the sun,  $\hat{r}_{\odot}=-\cos{\theta}\hat{x'}+\sin{\theta}\hat{y'}$, depicted in the inset on the upper left corner of Figure \ref{fig:schematic}.  We can solve for this angle by computing the dot product between the normal and the direction to the sun,

\begin{equation}
    \cos{\delta}=\frac{{\bf{\nabla}}f}{|{\bf{\nabla}}f|}\cdot\hat{r}_{\odot}=\frac{\sin{\psi}\sin{\theta}-\epsilon\cos{\psi}\cos{\theta}}{\sqrt{\sin^2{\psi}+\epsilon^2\cos^2{\psi}}}\,.
\end{equation}

For the surface-covering jet model, we assume that the outgassing is normal to the surface at each point.  The torque per unit mass, $\tau$, at a point on the surface, with the zenith angle modulation, is given by
\begin{equation}
\begin{split}
    \tau = -\bigg(b\,\alpha\frac{1-\epsilon^2}{\epsilon}\bigg)\\\frac{\cos{\psi}\sin{\psi}\big(\epsilon\cos{\psi}\cos{\theta}-\sin{\psi}\sin{\theta}\big)}{\big(\sin^2{\psi}+\epsilon^2\cos^2{\psi}\big)}\, .
\end{split}
\end{equation}
The total torque over the entire surface, $\Gamma$ is given by the line integral,
\begin{equation}
\begin{split}
    \Gamma =-\bigg(\frac{\alpha b^2(1-\epsilon^2)}{\epsilon^2}\frac{1}{\Delta \chi}\bigg)\\\oint_{\psi_-}^{\psi_-+\pi}
    \bigg(\frac{\cos{\psi}\sin{\psi}\big(\epsilon\cos{\psi}\cos{\theta}-\sin{\psi}\sin{\theta}\big)}{\sqrt{\sin^2{\psi}+\epsilon^2\cos^2{\psi}}}\bigg)d\psi\, ,
\end{split}
\end{equation}
where $\Delta \chi$ is the arclength of the line integral and the limits of the integral, $(\psi_+,\psi_-)$, define the endpoints of the illuminated surface. The arclength is given by

\begin{equation}
    \Delta \chi=\oint_{\psi_-}^{\psi_+}\bigg(b/\epsilon\sqrt{\sin{\psi}^2+\epsilon\cos{\psi}^2} \bigg)d\psi \, .
\end{equation}
Since we integrate over half of the ellipse, the arclength is given by 
\begin{equation}
    \Delta \chi(\epsilon)=\frac{2b}{\epsilon}\rm{E}(\sqrt{1-\epsilon^2})\, ,
\end{equation}
where $\rm{E}$ is the complete elliptic integral of the second kind. Note that $\rm{E}(e)$ is one quarter of the perimeter of an ellipse with eccentricity $e$. For an eccentricity of 0, or $\epsilon=1$, $\Delta \chi(\epsilon)=\frac{b}{\epsilon}\pi$, and as the eccentricity approaches 1, or the limit $\epsilon<<1$, $\Delta \chi(\epsilon)\rightarrow{}\frac{2b}{\epsilon}$. Evaluating the definite integral and simplifying, the total torque is,
\begin{equation}
\begin{split}
    \Gamma(\theta,\epsilon)
    =\bigg(\frac{\alpha b}{2\sqrt{1-\epsilon^2}}\frac{1}{\rm{E}(\sqrt{1-\epsilon^2})}\bigg)\\
    \bigg(\sin^{-1}{\bigg(\frac{\sqrt{1-\epsilon^2}\sin{\theta}}{\sqrt{\epsilon^2\cos^2{\theta}+\sin^2\theta}}\bigg)}\cos{\theta}-\\\epsilon\,\,\rm{sinh}^{-1}{\bigg(\frac{\sqrt{1-\epsilon^2}\cos{\theta}}{\sqrt{\epsilon^2\cos^2{\theta}+\sin^2\theta}}\bigg)}\sin{\theta}\bigg)\, ,
    \end{split}
\end{equation}
The moment of inertia per unit mass is  $I=a^2(1+\epsilon^2)/5$, and the associated hamiltonian is for the system is,
\begin{equation}
\begin{split}
\mathcal{H}=\frac{\Theta^2}{2}-\omega_0^2\bigg(\sin^{-1}\bigg(\frac{\sqrt{1-\epsilon^2}\sin{\theta} }{\sqrt{\epsilon^2\cos^2{\theta}+\sin ^2\theta}}\bigg)\sin{\theta}\\
+\, \epsilon\,\,\rm{sinh}^{-1}{\bigg(\frac{\sqrt{1-\epsilon^2}\cos{\theta}}{\sqrt{\epsilon^2\cos^2{\theta}+\sin^2\theta}}\bigg)}\cos{\theta}\bigg)
    \,,\end{split}\label{Hamiltonian}
\end{equation}
where $\Theta$ is the angular momentum per unit mass and  the frequency $\omega_0$ is defined as,
\begin{equation}
     \omega_0^2=\frac{5\alpha}{2a}\frac{\epsilon}{(1+\epsilon^2)(1-\epsilon^2)^\frac12}\frac{1}{\rm{E}(\sqrt{1-\epsilon^2})}\,.\label{frequency_omega0}
\end{equation}
\begin{figure*}[]
\begin{center}
\includegraphics[scale=0.8,angle=0]{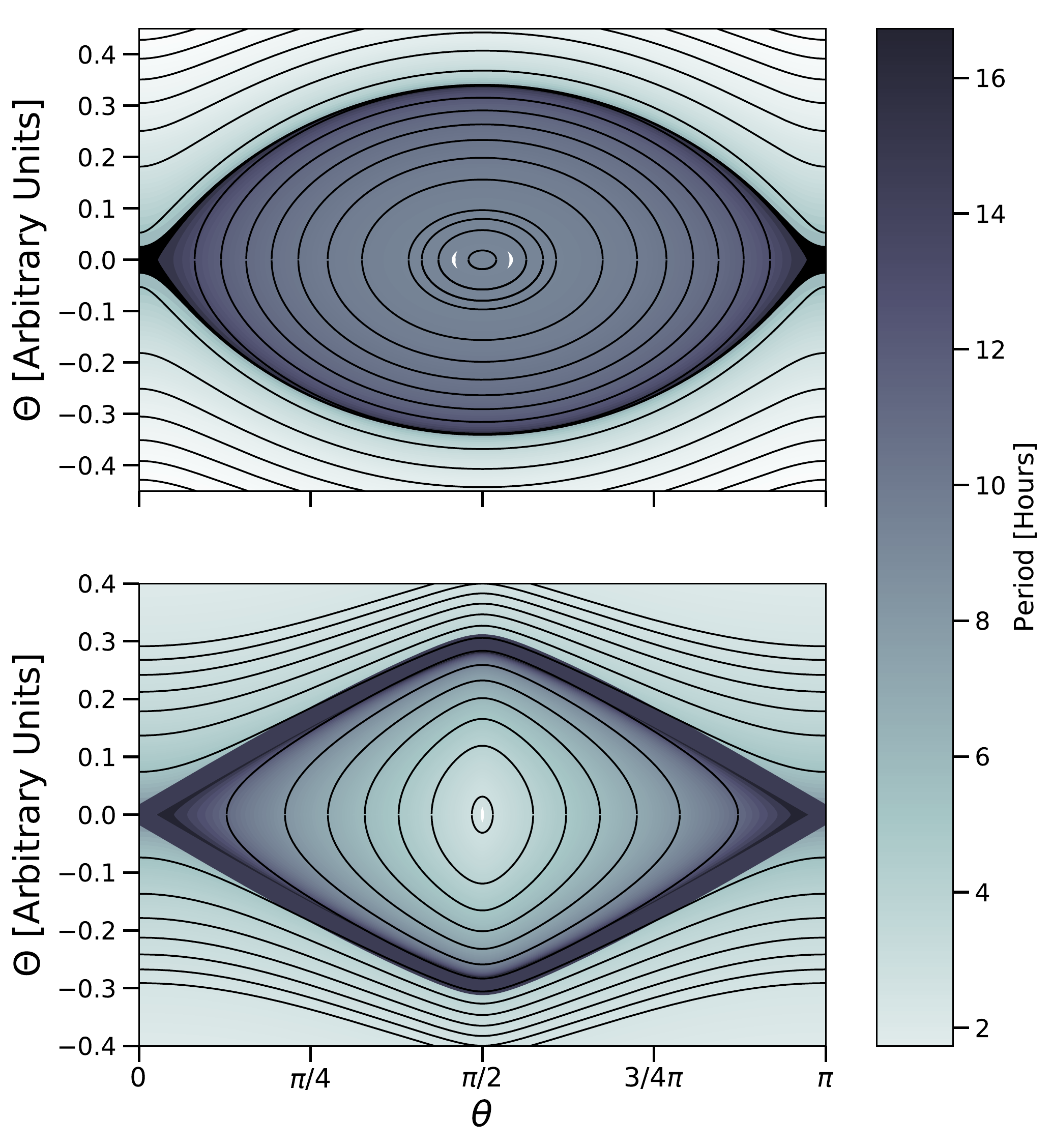}
\caption{Level sets of the Hamiltonians for the surface covering jet given by Equation \ref{Hamiltonian} (top panel) and single point jet given by Equation \ref{Hamiltonianpointjet} (bottom panel). The background color map shows the period for the librations or circulations. The overall character of the motion is not significantly changed by transitioning from a jet that vents only from the substellar point to one that emanates from an extended illuminated area.   \label{fig:hammy}}
\end{center}
\end{figure*}

\subsection{Three-Dimensional Tumbling}

The full 3D spin dynamics of the surface-covering jet model developed in the previous section are complex, and for our present purposes, we wish to merely establish whether such a model is likely to display secular spin-up. To evaluate this question, we first compare the  idealized 2D spin dynamics of the surface-covering jet  with the simpler substellar point-jet model developed in \cite{Seligman2019}. We then review the similarities that connect the dynamics of the analytic 2D and the numerical 3D single jet model, and we extend this approach to approximate the dynamics of of an ellipsoid with a surface-covering jet. As an approximation to the dynamics, we numerically simulate the 3D dynamical behavior of the surface-covering jet model by stochastically and rapidly changing the position of the jet. The calculations indicate that tumbling is permitted for the model and they also demonstrate that significant spin-up does not occur.

Within the framework of the single degree-of-freedom hamiltonian system developed in the previous section, it is evident that an ellipsoid with a surface-covering jet can exhibit pendulum-like oscillations about its spin axes, without undergoing significant spin-up. The top panel of Figure \ref{fig:hammy} shows the phase-space diagram for the Hamiltonian given by Equation \ref{Hamiltonian} for an ellipse of aspect ratio $6:1$. The background color corresponds to the period of the resulting oscillations, and the solid lines indicate the level sets of the Hamiltonian. The system exhibits a seperatrix denoting the transition between librating and circulating modes of oscillation. 

\cite{Seligman2019} demonstrated that the dynamics of a mobile jet emanating from the instantaneous substellar point also reduce to a one degree of freedom Hamiltonian, $\mathcal{H}_{ss}$, given by

\begin{equation}
\mathcal{H}_{ss}=\frac{\Theta^2}{2}-\omega_{ss}^2\frac{\sqrt{1+\epsilon^2-(\epsilon^2-1)\cos(2\theta)}}{\sqrt2(\epsilon^2-1)}
    \,,\label{Hamiltonianpointjet}
\end{equation}
where
\begin{equation}
     \omega_{ss}^2=\frac{5\alpha}{a}\frac{(1-\epsilon^2)}{(1+\epsilon^2)}\,.\label{frequency_omega0_pointjet}
\end{equation}
The phase space portrait and associated level sets for this Hamiltonian are shown in the bottom panel in Figure \ref{fig:hammy}, and are quite similar to those depicted in the top panel. 

The overall character of the motion is not significantly changed by transitioning from a jet that vents only from the substellar point to a series of jets that emanate from an extended illuminated area. This can be understood physically, without performing a detailed review of the equations presented in the previous subsection. The jets that emanate from the quadrant containing the substellar point (upper right) in Figure \ref{fig:schematic} dominate the dynamics and act similarly to a single jet acting at the substellar point. Moreover, the jets that span the two remaining quadrants that yield opposite torques are fairly weak, and therefore do not conspicuously alter the dynamics. 

Comparison of the equations governing the point jet dynamics (Equations \ref{Hamiltonianpointjet}-\ref{frequency_omega0_pointjet}) with those governing the surface-covering jet dynamics (Equations \ref{Hamiltonian}-\ref{frequency_omega0}) shows the strong similarity between the two systems. Most importantly, the frequency of the oscillations induced by the surface-covering jet are comparable in magnitude to those of the single jet,  $\omega_0\sim \sqrt{\epsilon/2}\, \omega_{ss}$. Moreover, the phase space structures of both systems are qualitatively similar, as seen by comparing the two phase space portraits presented in Figure \ref{fig:hammy}. The most striking difference between the two systems is the structure of the level sets of the resulting Hamiltonians, particularly in the vicinity of the central $\theta=\pi/2$ jet venting angle. The single point jet exhibits level sets with sharp central peaks, while the level sets for the surface covering jet show a shallower evolution in this region. This difference arises because the region close to $\theta=\pi/2$ corresponds to the orientation where the long axis of `Oumuamua is facing the Sun. This orientation corresponds to a minimum of the potential energy associated with the jet, and a maximum of kinetic energy in the rotation. An ellipsoid with a single point jet quickly passes through this potential energy minimum, like a pendulum, which manifests as the sharp peak. In contrast, the surface covering jet damps the motion through this central region, as the jets emanating from both sides of the axes of symmetry supress the net torque. Nonetheless, `Oumuamua's light curve places it in the circulating regime of phase space, where these differences are minimal. 
\begin{figure*}[]
\begin{center}
\includegraphics[scale=0.5,angle=0]{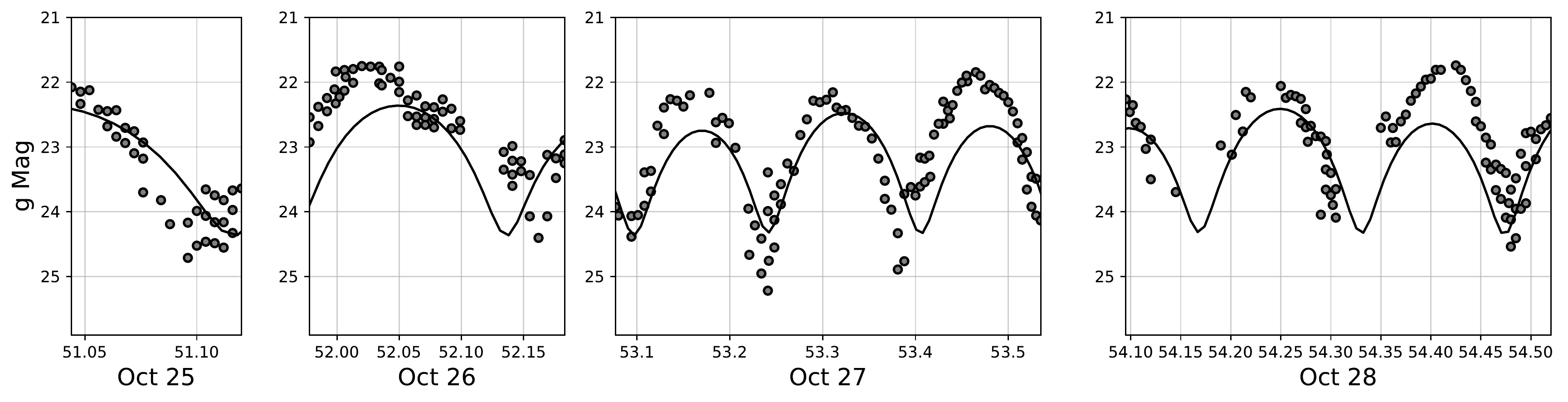}
\caption{The synthetic light curve  of `Oumuamua for the 3D dynamical evolution of the surface covering jet. Grey points show the real data that was taken for `Oumuamua (reproduced from \citet{Belton2018}), and the black line shows the synthetic light curve. Although we computed the spin evolution for the entirety of the trajectory through the inner Solar System, we only show the period of time between October 25-28, when the highest cadence data was available.}\label{fig:lightcurve}
\end{center}
\end{figure*}

\begin{figure}[]
\begin{center}
\includegraphics[scale=0.5,angle=0]{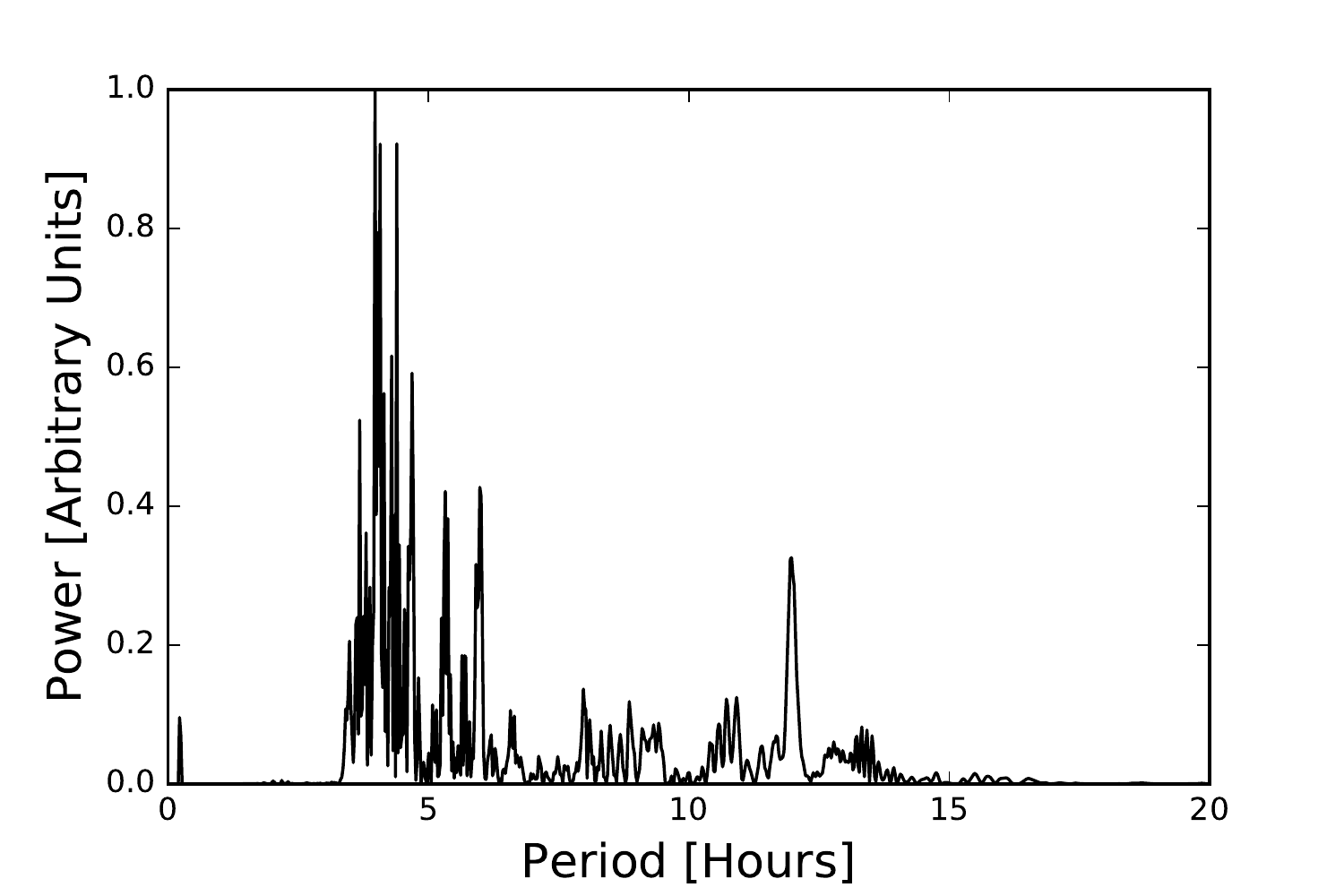}
\caption{The periodogram of the synthetic light curve  presented in Figure \ref{fig:lightcurve} for the 3D dynamical evolution of the surface covering jet.}\label{fig:power_spectrum}
\end{center}
\end{figure}

\cite{Seligman2019} computed the full 3D tumbling of an ellipsoid experiencing continuous torque from a single point jet, including the effects of thermal inertia and stochastic outgassing. They found that the 2D idealized dynamics provided a good approximation to the full 3D motion. Specifically, the 3D evolution under the action of the jet produced no secular increase in angular momentum, in agreement with the 2D approximation. Moreover, the synthetic light curves generated from the 3D motion broadly reproduced the observed power spectrum and the observed range of brightness fluctuations in the real data. In order to verify that our foregoing 2D analytic solution for the surface covering jet provides a similarly valid approximation to the full 3D dynamics, we compute the dynamical evolution of the oblate ellipsoid model under the influence of a point jet whose position and magnitude is stochastically forced. In order to stochastically force the position and magnitude of the jet, we applied random variations in magnitude of functional form
\begin{equation}
    \alpha(t+\delta t)=\alpha (t) e^{-\delta t/\tau}+\Xi\sqrt{1-e^{-2\delta t/\tau}}
    \, ,\label{stocastic_forcing}
\end{equation}
\citep{Rein2010}, where $\tau\sim 1\, \rm{hr}$ is the auto-correlation timescale and $\Xi$ is a random variable with normal distribution to the position of the jet. The randomly perturbed point jet serves as an approximation to the surface covering jet, and is not directly analagous. Computing the full 3D dynamics of the entire surface covering jet becomes an expensive  computational problem. Formally, one would have to  apply torques at an infinite number of points on the surface to simulate the correct dynamics. As the 2D Hamiltonian well-approximates the dynamics of a general point jet, our Hamiltonian analysis should also provide a valid approximation to the dynamics.

In Figure \ref{fig:lightcurve}, we show the synthetic light curve generated from the simulation, compared to the real data that was taken for `Oumuamua. Although we computed the light curve for the entirety of the trajectory through the inner Solar System, we only show the period of time corresponding to the days in October 2017 when high cadence observations were taken. We produced the synthetic light curve using open source ray-tracing software\footnote{\url{http://www.povray.org/}}, using a similar methodology presented in \citet{Seligman2019}. We perform the ray-tracing by placing an ellipsoid of the best fit axis ratio, and  the camera, corresponding to the observer on Earth and light source, corresponding to the Sun, in their  orbital locations. We then calculate the illumination based on the diffuse reflectivity of the object, and  sum the brightness to generate an unresolved flux. As the ellipsoid rotates under the action of the stochastically forced jet, we compute the flux at each point in time. It is important to note that  because  the dynamical model employed here is very idealized and only an approximation to the surface covering jet, we did not attempt to fit the light curve that was observed for `Oumuamua, as was done extensively by \citet{Mashchenko2019}. We merely want to demonstrate that  a jet that covers the  illuminated surface can reproduce the qualitative features of the light curve. While our synthetic light curve is certainly not an exact fit to the data, it provides a modest approximation to the stability in the periodicity.

In Figure \ref{fig:power_spectrum}, we show the resulting periodogram of the synthetic light curve shown in Figure \ref{fig:lightcurve} for the entire trajectory. This power spectrum may be compared to the four periodograms presented in Figure 3 in \citet{Seligman2019}. Evidently, the light curve for the extended jet model exhibits no secular change in spin period, nor does the light curve for the point jet model. The extended jet model is designed to mimic the effects of high surface covering outgassing, while the single jet model is more representative of low surface covering outgassing (such as the H$_2$ outgassing scenario). In reality, these two models are the two extreme cases, and `Oumuamua's outgassing was likely in between the two regimes. The exact surface covering fraction, which is unknown, would determine which regime the tumbling dynamics most mimicked. Nevertheless, the lack of spin-up is consistent with outgassing in either case.

\subsection{Change in Spin State from Radiation Pressure}
 While the purpose of this section is to examine the spin dynamics of jets that emanate from the entire illuminated surface, this formalism is readily applied to evaluate the spin state from radiation pressure. In this section, we demonstrate that radiation pressure will not change the angular momentum of an ellipsoidal body. This provides supportive evidence for  theories which  invoke  radiation pressure for the source of the anomalous acceleration, such as those presented in  \citet{MoroMartin2019}, \citet{luu2020oumuamua}, and \citet{bialy2018could}.

It is straightforward to show that the  torque from radiation pressure on an ellipsoidal body is zero. At any point on the surface, the force from radiation pressure is parallel to the axis connecting the Sun and the center of mass. The total torque, $\Gamma_{rad}$, is
\begin{equation}
\begin{split}
    \Gamma_{rad} =-\bigg(\frac{\alpha b}{\epsilon}\frac{1}{\Delta \chi}\bigg)\\\oint_{\psi_-}^{\psi_-+\pi}\big(\sin{\psi}\sin{\theta}-\epsilon\cos{\psi}\cos{\theta}\big)\\\,\,
    \big(a\cos{\psi}\sin{\theta}+b\sin{\psi}\cos{\theta}\big)\,\,d\psi\, .
\end{split}
\end{equation}

The points ($\psi_-,\psi_-+\pi$) are antipodal and therefore $\Gamma_{rad}=0\,\,\forall\theta\in(0,2\pi)$   $\forall \psi_-\in(0,2\pi).$

\section{Energetic Constraints on Carbon Monoxide as the Accelerant}\label{sec:energetics}

\begin{figure}[]
\begin{center}
\includegraphics[scale=0.3,angle=0]{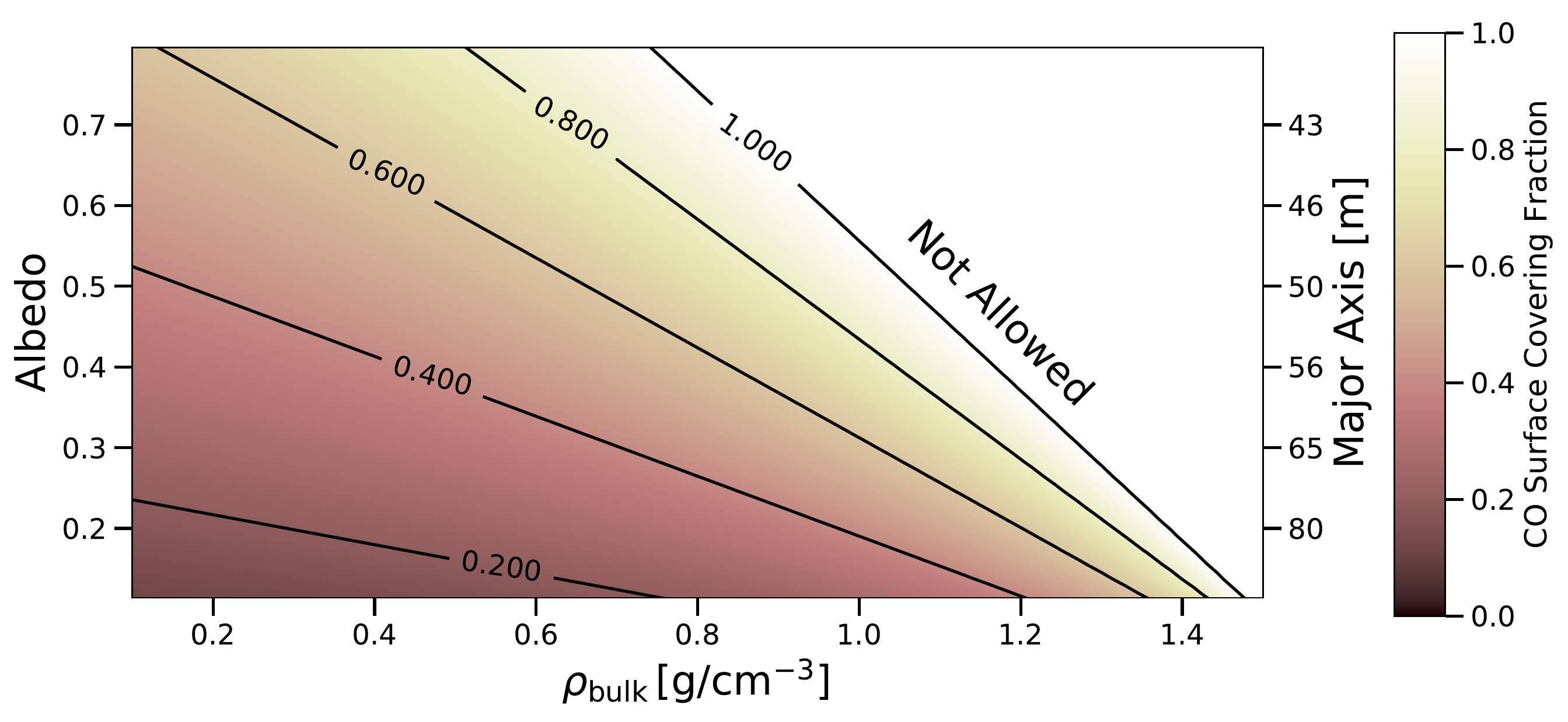}
\caption{The fraction of the illuminated surface of `Oumuamua that would have needed to be covered in carbon monoxide ice to power its non-gravitational acceleration. We show the fraction for a plausible range of bulk densities and surface albedo (and corresponding size depicted on the secondary y-axis). The color scale shows the fraction of the illuminated surface required, and the contours are plotted up until 100$\%$ of the surface.  \label{fig:CO_fraction}}
\end{center}
\end{figure}
\citet{Seligman2020} (hereafter SL20) demonstrated that the time-dependent flux of solar irradiance provided a constraint on the composition of `Oumuamua. The energy input from solar radiation, $E_{\mathrm{tot}}=4.5\times10^{13}\,{\rm erg\,cm^{-2}}$ (integrated over the two years surrounding perihelion), must provide sufficient energy to power the non-gravitational acceleration that was observed. This energy provides  both the latent heat of sublimation and the kinetic energy of each gas molecule (see Equation 1 in SL20). Table 1 of SL20 demonstrated that H$_2$, N$_2$, Ne and Ar were permissible accelerants, and ruled-out other common cometary volatiles as the dominant accelerant.

In this section, we demonstrate that CO is also energetically viable. This species was not considered in SL20 because of the CO limits from the  \textit{Spitzer} non-detection, which we address  in \S\ref{sec:supportive_evidence_astrometry} and \S\ref{sec:supportive_evidence_galactic_reservoir}. SL20 quantified the viability of an accelerant by the fraction of the illuminated surface that is required to be actively outgassing in order to explain the non-gravitational acceleration.  Equation 6 from SL20 may be used to calculate the  surface covering fraction for a given volatile species, $f$, and reads,

\begin{equation}
\begin{split}
f=\bigg(4(\Delta H/N_A+\gamma k T_S)\rho_{\rm{bulk}} c\vert \alpha(t) \vert \bigg)/ \\
\bigg(((1-p)Q(t)-\epsilon \sigma T_S^4)(9\mu m_{\rm u}\gamma k T_S)^{\frac{1}{2}}\\\xi(1+(\frac{c^2}{ea^2})\tanh^{-1}e)\bigg)\,.  \label{eqn:surface_fraction}
\end{split}
\end{equation}
In this equation,  $Q(t)$ is the local solar irradiance, $\alpha$ is the magnitude of the non-gravitational acceleration, $\epsilon$ is the surface emissivity, $\Delta H$ is the sublimation enthalpy of the ice, $T_S$ is the sublimation temperature, $\gamma$ is the adiabatic index of the escaping vapor,  $p$ is the surface albedo,  $\mu m_{\rm u}$ is the mass per molecule of a species and $\rho_{\rm{bulk}}$ is the bulk density. The parameter, $\xi$, was defined in SL20 and  denotes the average projected surface area of the body divided by its total surface area. When isotropically averaged over all viewing angles, $\xi=1/4$ for any convex body \citep{Meltzer1949}. Using the digitized light curve from 5 Oct. 2017 through 1 Nov. 2017 \citep{Belton2018}, SL20 calculated  $\xi_{oblate}\sim0.2$. They assumed the  disk-like $a$:$a$:$c\sim6$:$6$:$1$ most-probable model given by \citet{Mashchenko2019}, which has physical dimensions 115 m x 111 m x 19 m. In Equation \ref{eqn:surface_fraction}, $e=\sqrt{1-c^2/a^2}$ is the ellipsoid's eccentricity. The factor of $((1-p)Q(t)-\epsilon \sigma T_S^4)/(\Delta H/N_{A}+\gamma kT_S)$ corresponds to the flux of sublimated molecules leaving a directly illuminated patch of surface ice, and is written out in full form as defined in  Equation 1 of SL20 and in Equation \ref{eq:calN} of this paper.


In Figure \ref{fig:CO_fraction} we show the required surface covering fraction for CO ice to satisfy the non-gravitational acceleration, as a function of  albedo and bulk density. For carbon monoxide, $\Delta H=8.1 \,{\rm kJ\,mol}^{-1}$ and  $\rm{T_{S}}=60\,$K\footnote{NIST Chemistry WebBook for CO at \href{https://webbook.nist.gov/cgi/cbook.cgi?ID=C630080&Mask=4}{this link.}}. For a  plausible range of bulk density and albedo, carbon monoxide is a viable accelerant. It is important to note that in order to calculate the required surface covering fraction as a function of albedo, each axis of the ellipsoid  scales by a factor $\sim \sqrt{0.1/p}$. 

\section{Challenges for the CO Hypothesis}\label{sec:challenges}

\subsection{CO Outgassing Constraints from \textit{Spitzer}}

The conclusions of \citet{Trilling2018} were based on \textit{Spitzer} exposures which were cleaned of sidereal sources and shift-stacked along `Oumuamua's calculated trajectory (which included the non-gravitational acceleration). The resulting image did not reveal the significant detection of `Oumuamua in the IRAC Channel 2 band ($1\mu$m-wide, centered on $4.5\mu$m). The image's noise level of 0.1$\mu$Jy (1$\sigma$) was verified on the basis of injection/recovery tests. \citet{Trilling2018} placed stringent limits on the production rates $Q$ of dust, CO, and CO$_2$ based on their expected line emission and the observed \textit{Spitzer} flux density limit.

\citet{Trilling2018} used the \citet{Haser1957} model for cometary coma to describe the CO$_2$ number density profile,
\begin{equation}
    n(r) = \frac{Q({\rm CO}_2)}{4\pi r^2 v}\exp(-\frac{r}{\tau v})\, ,
\end{equation}
for radial outflow velocity, $v$, and CO$_2$ photodissociation lifetime, $\tau$. 
The column density is related to the production rate and flux density $F$ as,
\begin{equation}
    N(r) = \frac{4\pi F \cdot 10^{-9}}{g_{\nu_3} Q({\rm CO}_2) hc \pi r^2/\lambda}\Delta^2\, ,
\end{equation}
where $\Delta$ represents the distance to the observer, $g$ represents the fluorescence efficiency (subscript $\nu_3$ denotes the line transition), and $r$ is the distance from the nucleus. \citet{Trilling2018} calculated that $Q({\rm CO}_2) < 9\times10^{22}$ s$^{-1}$ ($3\sigma$). The ratio of
fluorescence efficiencies for the CO(1-0) transition and the CO$_2$ $\nu_3$ transition is $\sim 0.1$, assuming $g_{1-0} = 2.46\times10^{-4}$ and $g_{\nu_3} = 2.86\times10^{-3}$ \citep{Crovisier1983}. The implied upper-limit on CO production is  $Q(\rm CO)\sim9 \times 10^{23}$ molecules s$^{-1}$, as opposed to $Q(\rm CO) \sim 9 \times 10^{21}$ molecules s$^{-1}$ quoted in \citet{Trilling2018}, which was obtained (incorrectly)\footnote{Note that the discrepancy stems from a typo in applying the fluorescence efficiencies.} by dividing the CO$_2$ gas production limit by a factor of 10.  Accounting for the energetics of the non-gravitational acceleration via Equation 1 in SL20 gives

\begin{equation}\label{eq:calN}
{\cal N}=\frac{(1-p)Q(t)-\epsilon \sigma T_S^4}{\Delta H/N_{A}+\gamma kT_S}\, ,
\end{equation}
and we calculate that the outgassing rate per unit surface area is,

\begin{equation}
    Q({\rm CO})=2.1\times 10^{18} {\rm s^{-1} \, cm^{-2}} \bigg(\frac{a}{ 110 {\rm m}}\bigg) ^2\bigg(\frac{r}{2\mathrm{au}}\bigg)^{-2}\,.\label{eqn:qCO}
\end{equation}

It is important to note that this is still inconsistent with the revised \textit{Spitzer}  estimates by a factor of $\sim 100-1000$ based on the assumptions of the albedo and size of `Oumuamua. For a major axis of 115m and albedo of 0.1 at 2au, $Q(\rm CO)\sim4.5 \times 10^{26}$ molecules s$^{-1}$, and for a major axis of 50m and albedo of 0.5, $Q(\rm CO)\sim9.4 \times 10^{25}$ molecules s$^{-1}$. Therefore, we conclude that if `Oumuamua's non-gravitational acceleration was powered by the sublimation of CO, then the CO production must have significantly lowered at some point prior to the \textit{Spitzer} observations. This is seemingly problematic for the validity of the CO hypothesis, since it is unlikely that the activity would stop before we had the chance to observe it. However, cometary outgassing is known to be sporadic, which may be attributed to seasonal effects \citep{Kim2020} or uneven volatile layers in the interior (see \S \ref{sec:solar_system_CO} and the references therein). Moreover, CO outgassing would not have been observable prior to the \textit{Spitzer} observations, since `Oumuamua was  discovered post perihelion, 5 days after its closest  approach to the Earth and high quality observations were only obtainable for $\sim 1$ week. 

\subsection{Other Issues with CO Ice}

While outgassing powered by CO would explain `Oumuamua's bulk composition with commonly-observed solar system volatiles, this hypothesis still does not perfectly describe the interstellar object's properties. Despite the upward revision in allowable CO outgassing, the sublimation rate required to satisfy the non-gravitational acceleration at $2\,\text{au}$ is still in tension with the \textit{Spitzer} non-detection, as we demonstrated in the previous subsection. The only plausible way that `Oumuamua was CO-rich, is if the outgassing was  sporadic, and the nucleus was inactive when viewed by \textit{Spitzer}. \cite{Micheli2018} reported non-gravitational acceleration in approximately one-month pieces of the interstellar object's trajectory, but the astrometric data does not allow for significant detection with further subdivision.

Therefore, the CO hypothesis requires a degree of fine-tuning of `Oumuamua's bulk composition and activity to simultaneously satisfy the non-gravitational acceleration and \textit{Spitzer} observational constraints. Furthermore, the lack of any visible coma at any point in the trajectory presents issues if `Oumuamua is considered a direct, but smaller analogue of the CO-rich Borisov \citep{Bodewits2020, Cordiner2020}.  Upper limits on the production rate of micron sized dust particles by `Oumuamua were estimated to be  $Q({\rm dust})<2\times 10^{-4}$ kg  s$^{-1}$ \citep{Jewitt2017} and  $Q({\rm dust}) <1.7\times10^{-3} {\rm kg \,s}^{-1}$  at a distance of 1.4 au \citep{Meech2017}. However, \citet{Micheli2018} argued that cometary outgassing was still the most likely explanation for the non-gravitational acceleration, despite the stringent limits on dust production. They demonstrated that if the dust component was dominated by  grains larger than a few hundred micrometers to millimeters, they would not have been detected at optical wavelengths. There were no observations made that could have sensitively detected large dust, given the short observing window and faintness of the object.  The solar system comet 2P/Encke exhibited a dearth of small dust grains close to its perihelion passage, based on the low continuum level measured relative to its emissions \citep{Fink2009}. In this study, the authors argued that this feature was ``probably caused by an exhaustion of small dust grains after its many orbits close to the sun, so that only large grains with smaller total scattering cross section remain''. Subsurface layers of comets are expected to be more enriched with larger dust grains \citep{Laufer2005,Micheli2018}, and since  `Oumuamua was detected post-perihelion, it is likely that it had already lost a fraction of its surface \citep{Seligman2019,Seligman2020,desch20211i,jackson20211i}. It is possible that due to extensive processing in interstellar space, small refractory particles on interstellar objects will be liberated from the surface \citep{Stern1987,Stern1990}. Specifically, \citet{Stern1990} demonstrated that the interstellar medium  efficiently removes  small dust grains from Oort cloud comets by drag effects, which also applies to interstellar objects. Interestingly, this effect is particularly effective during the passage through supernovae remnants, and for a detailed analysis we refer the reader to Section 6 of \citet{Stern1990}. Moreover, there is evidence that CO is not substantially incorporated into the dust component of the cometary nucleus of 29P (see \S \ref{sec:solar_system_CO}). 

If the CO hypothesis is correct, then `Oumuamua's temporal and spatial correlation with the Carina and Columba associations \citep{mamajek2017, Hallatt202} would most likely be coincidental. The metal budget of the protoplanetary disks born in these regions cannot produce a local population of interstellar objects  as large as the one inferred by \cite{Do2018}, so observing one of these primordial planetesimals in Pan-STARRS would be improbable. Nevertheless, the inferred paucity of scattering events during `Oumuamua's interstellar journey suggests a young age. Without a destructive mechanism to remove older members of the CO-rich interstellar object population, `Oumuamua would have been selected from among the youngest objects. While not impossible, this event is unlikely.

\section{Astrometric Constraints on CO Outgassing}\label{sec:supportive_evidence_astrometry}

Here, we examine whether astrometric constraints from `Oumuamua's trajectory can immediately preclude the sporadic activity scenario required to reconcile CO outgassing with the \textit{Spitzer} non-detection. Specifically, we search for evidence of non-gravitational acceleration in relevant subsets of the observational arc. While \textit{Spitzer} eliminates the possibility for CO-driven activity during the observational window, the majority of the astrometry for `Oumuamua was obtained a prior to this non-detection.

\cite{Micheli2018} determined that `Oumuamua experienced significant non-gravitational acceleration and fit continuous $1/r$ or $1/r^{2}$ heliocentric distance functions to the astrometric data for the non-Keplerian component. We corroborate this detection of a non-ballistic trajectory and demonstrate that the observations also cannot disallow variable outgassing. Although this interpretation is {\it ad hoc} and entirely retrospective, it merits investigation given the complications with exotic objects and that the activity levels of Solar System comets are known to change over time.

\subsection{Validating the Non-Gravitational Acceleration}

We use \texttt{Find\_Orb}\footnote{\url{https://www.projectpluto.com/find\_orb.htm}}, an open-source package designed to determine orbital elements from a set of ephemerides. Data were obtained from the Minor Planet Center\footnote{\url{https://www.minorplanetcenter.net/}, retrieved 05 Jan. 2021} (MPC). We consider the same 207 observations with the same uncertainties on right ascension and declination as \cite{Micheli2018}. The temporal uncertainty was set to $1\,\text{s}$ for all observations. 

Fitting `Oumuamua's entire trajectory with the model of \cite{Marsden1973}, we find radial non-gravitational parameter $A_{\text{1}} = (2.37 \pm 0.09)\,\times10^{-7}\,\text{au/day}^{2}$ and no significant $A_{\text{2}}$. For comparison, \cite{Micheli2018} found $A_{\text{1}} = (2.45 \pm 0.08)\,\times10^{-7}\,\text{au/day}^{2}$ with a proprietary JPL orbit fitting code. In summary, these two orbital fits recover non-gravitational acceleration at $26\sigma$ and $30\sigma$ significance, respectively, and are consistent within their uncertainties.

Interestingly, the magnitude of $A_{\text{1}}$ for `Oumuamua places it around the 95$^{\text{th}}$ percentile of the approximately 200 objects in the JPL Small-Body Database\footnote{\url{https://ssd.jpl.nasa.gov/sbdb.cgi}} with a determined value \citep{Micheli2018}. The specific volatile compositions and bulk densities of most small bodies with higher $A_{\text{1}}$ than `Oumuamua are unknown, as they were not observed as intensely.

\cite{katz2019evidence} points to a reduction in the scatter of residuals between the gravitational and non-gravitational models in \cite{Micheli2018} as requiring further investigation. However, this difference in behavior of the residuals can be explained by the fact that these data were collected from different telescopes with different levels of precision. Specifically, four observatories obtained astrometry for `Oumuamua on 19 Oct. 2017, which \cite{katz2019evidence} identifies as a potentially problematic date. For example, the ESA Optical Ground Station (OGS) Telescope (MPC code J04), is assigned uncertainty $\sigma = 0.2"$ for both right ascension and declination, while $\sigma = 3.0"$ is reported for the Klet' Observatory (MPC code 246).

Adding a non-gravitational parameter to `Oumuamua's trajectory fit modifies a given predicted position in the sky by the same absolute amount, irrespective of the point's assigned uncertainty. Figure 2 in \cite{Micheli2018}, which is flagged as concerning by \cite{katz2019evidence}, shows the relative residuals normalized by $\sigma$ for each observatory. Thus, the effect of observatory-dependent denominators $\sigma$ manifests itself as the collapse in the scatter of residuals identified by \cite{katz2019evidence}.

\begin{figure}[]
\begin{center}
\includegraphics[scale=0.5,angle=0]{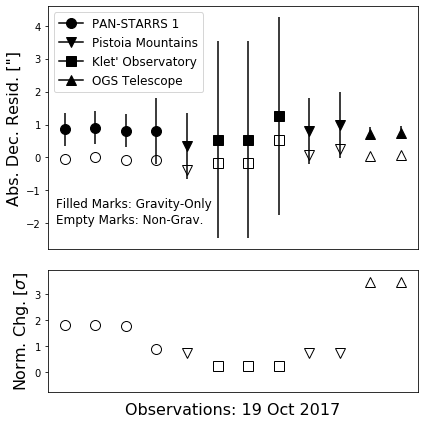}
\caption{Declination residuals for the observations of `Oumuamua on 19 Oct. 2017 from the trajectory fits of \cite{Micheli2018}. Points are equally-spaced on the x-axis but are displayed in temporal order from left-to-right. Panel (a) shows the absolute residuals in arcseconds for the gravity-only model $\delta_{\text{g}}$ (filled marks) and non-gravitational model $\delta_{\text{n}}$ (empty marks). Error bars representing the uncertainty $\sigma_{\text{d}}$ are shown on the gravity-only model. Panel (b) shows the normalized change in residuals, $(\delta_{\text{g}} - \delta_{\text{n}}) / \sigma_{\text{d}}$, demonstrating that the collapse in scatter discussed by \cite{katz2019evidence} comes only from the different uncertainties of the observatories.}\label{fig:katz}
\end{center}
\end{figure}

We illustrate the change in residuals between the best-fit ballistic and non-Keplerian trajectories in Figure \ref{fig:katz}. The top panel shows the absolute declination residuals for the observations taken on 19 Oct. 2017. Because all of these data were obtained within the same night, a change in model affects them all similarly. As expected, the difference between the two models for a given observation is small.  On the bottom panel, we show the change in residual between the models, normalized by the uncertainty for each observation. In relative terms, points from the OGS telescope move the most simply because they have the lowest uncertainty.

\subsection{Interrogating Subsets of the Observational Arc}

In addition to finding a consistent fit with continuous non-gravitational acceleration, \citet{Micheli2018} ruled-out a single impulsive event due to significant non-Keplerian dynamics at the beginning and end of the observational arc. However, \cite{Micheli2018} did not consider sporadic outgassing, a scenario between these two extremes and the only possible mechanism to reconcile `Oumuamua's CO-driven non-gravitational acceleration with the \textit{Spitzer} non-detection.

First, we examine whether significant non-gravitational acceleration can be identified without the astrometric data gathered around the time of the \textit{Spitzer} exposures. Specifically, we remove points from 21 Nov. 2017, the day of the \textit{Spitzer} observation. While \textit{Spitzer} itself obtained no astrometry given its non-detection of `Oumuamua, the Las Campanas Observatory (MPC code 304) provided contemporaneous ground-based data. Using the 190 remaining points in the trajectory, we find the same value: $A_{\text{1}} = (2.37 \pm 0.09) \,\times10^{-7}\,\text{au/day}^{2}$. Therefore we infer that the astrometric data around the time of the \textit{Spitzer} non-detection is not required to confidently-determine the overall fit of the trajectory.

Next, we attempted to find signatures of non-gravitational acceleration specifically in the time surrounding the \textit{Spitzer} observations. We investigated the subset of data from 15 Nov. 2017 through 23 Nov. 2017, within a week of the attempted \textit{Spitzer} observation. With the same \texttt{Find\_Orb} routine, we find no significant deviation from a Keplerain trajectory: $A_{\text{1}} = (-2.01 \pm 4.0)\,\times10^{-6}\,\text{au/day}^{2}$.

Confidently-determining $A_{\text{1}}$ in `Oumuamua's trajectory seems to require an arc of around one month. Therefore, the astrometric data is not precise enough to verify that `Oumuamua was accelerating non-gravitationally specifically while the \textit{Spitzer} images were being collected. Thus, the observations do not appear to rule-out the variable activity scenario required to reconcile CO outgassing with the \textit{Spitzer} non-detection.

\section{ Galactic Reservoir of CO-Rich Objects}\label{sec:supportive_evidence_galactic_reservoir}

Estimates of the galactic number density, $n_{\text{iso}}$, of interstellar objects are broadly consistent with planetary formation theory if the population consists of objects ejected from extrasolar systems. \cite{Do2018} found that the detection of `Oumuamua implied $n_{\text{iso}} \sim 0.2\,\text{au}^{-3}$ via a detailed analysis of the Pan-STARRS survey volume. This calculation refined the order-of-magnitude estimates presented in  \citet{Trilling2017}, \citet{Laughlin2017}, and \citet{Jewitt2017}. 

To assess the feasibility of populating the galaxy with sufficient `Oumuamua-like objects, we draw a comparison with estimates of the material ejected from the solar system. In models of the dynamical evolution of the early solar system, migration of the giant planets populated the Kuiper Belt and Oort Cloud, and for a recent review of different formation models we refer the reader to \citet{Nesvorny2018}. An example formation history, ``Nice model," suggests that a transient dynamical instability of the giant planets generated $\sim 30\text{\mearth}$ of material in interstellar comets \citep{Tsiganis2005,Levison2008}. Planetesimal-driven migration \citep{Hahn1999,Gomes2004} of the giant planets would also generate an order-of-magnitude similar amount of mass in ejected planetesimals. If `Oumaumua was CO-rich, then it may indicate that a significant fraction of material ejected from extrasolar systems is also CO-rich. We estimate an upper limit of order $10\,\text{\mearth}$ of carbon and oxygen could be incorporated into interstellar comets, much of it as volatiles such as CO, based on the protosolar abundances of these elements \citep{draine2011book}.

To quantify the resulting number density of interstellar objects, we assume that `Oumuamua is representative of a class of CO-rich interstellar objects with typical cometary density born in protoplanetary disks forming at approximately the Milky Way's star-formation rate of $\sim3\,\rm{yr}^{-1}$ \citep{Shu1987}. While the survival timescales of interstellar objects depend on their composition and the environments through which they pass, we adopt a fiducial lifetime of order $10\,\text{Gyr}$. This value corresponds to roughly double the lower bound from the survival of processed small bodies in the Oort Cloud. By using this conservative lifetime, we demand a larger available CO mass reservoir than would be required for a longer timescale. Taking these objects to be uniformly distributed in the approximately $10^{67}\,\text{cm}^{3}$ in the galaxy, we find $n_{\text{iso}} \sim 0.6\,\text{au}^{-3}$. 

This value is in general agreement with \cite{Do2018}. However, the Pan-STARRS survey time has roughly doubled since this observational constraint was placed on the interstellar object population. Nominally, this additional search volume without another detection should revise downward the estimated number density by a factor of a few. Nonetheless, this slight revision does not affect the reconciliation of the possible CO mass reservoir with the inferred occurrence of `Oumuamua-like objects at the order-of-magnitude level. Furthermore, the statistical uncertainties stemming from a single detection are far larger than any necessary refinements to this analysis attributed to a longer Pan-STARRS campaign. Moreover, these interstellar objects likely draw from a larger mass reservoir by incorporating elements other than carbon and oxygen.

We have taken the lifetime of `Oumuamua-sized CO objects to be of order $10\,\text{Gyr}$, although the survival of this ice in the ISM is relatively unconstrained. Within the assumed timescale, `Oumuamua would be among the most recently-formed of its peers as inferred from its kinematic age of less than $100\,\text{Myr}$ \citep{Hallatt202}. However, the solar system's current position in the Milky Way is close to the galactic midplane, leading to a higher probability of encountering younger objects (Hsieh et al. submitted). 

A confounding detail is that \citet{MoroMartin2018i} demonstrated that `Oumuamua was unlikely to come from an isotropically distributed population of icy planetesimals in the galaxy, by comparing the expected mass budget of ejected planetesimals around extrasolar systems to the  mass budget of interstellar objects inferred from the detection of `Oumuamua. The inferred mass budget relies critically on the size, albedo and bulk density of `Oumuamua, and was calculated assuming a  range of  size-frequency distributions for the ejected planetesimals, which were based on observed solar system distributions. The study concluded  that to rectify this inconsistency, `Oumuamua likely originated from a planetesimal disk of a young nearby star whose remnants were highly anisotropic. This conclusion is also consistent with `Oumuamua's previously-described kinematics of `Oumuamua. Given the range of albedos (and corresponding sizes) and bulk densities that are allowable for `Oumuamua with CO as the accelerant shown in Figure \ref{fig:CO_fraction}, it is possible that `Oumuamua had a smaller size and/or bulk density than those considered by \citet{MoroMartin2018i}, which could also rectify the inconsistency.  

\section{CO in Solar System Comets}\label{sec:solar_system_CO}

CO is a common volatile in the Solar System. Objects such as active Centaurs located in regions too cold for substantial H$_2$O ice sublimation are believed to have activity that is driven by other substances such as CO \citep{Womack2017}. For example, \citet{Jewitt2009} examined 23 Centaurs and found 9 to be active. Moreover, the study presented compelling evidence that some of the objects' activity was driven by the release of trapped gases, included CO, during the conversion of amorphous into crystalline ice. As another example of CO-driven activity, \citet{Bauer2015} found that 40 objects in a sample of 163 comets from the Wide-Field Infrared Survey Explorer (WISE) had excess flux in a band centered at 4.5 micron. Since both CO and CO2 are encapsulated in this broad band, Table 4 of their study lists the approximate production rates for each object as inferred by the fluorescence efficiencies. Figure 8 of their paper shows that the objects have production rates $Q({\rm CO}_2)$ between $10^{25}$ and $10^{28}$ molecules s$^{-1}$, corresponding to $Q({\rm CO})$ between $10^{24}$ and $10^{27}$ molecules s$^{-1}$. These values are broadly consistent with the levels that would be required for `Oumuamua. However, typical objects that are observed at far distances are believed to have other volatile species such as H$_{2}$O present, but simply inactive.

It is also not unprecedented to have non-detections of CO in comets in the Solar System. Images of the comet C/1995 O1 Hale-Bopp with two telescopes at the Observatorio del Teide failed to reveal the expected CO emission, suggesting that the CO production was very low or from a very extended source \citep{Santos-Sanz1997}. In the \textit{Akari} near-infrared survey of 18 comets, CO was only detected in 3 comets (29P/SW1, C/2006 W3 (Christensen) and C/2008 Q3 (Garradd)) and the ratio of CO/CO$_2$ was less than unity for all of the comets \citep{Ootsubo2012}. \citet{Drahus2017} presented significant non-detections and upper limits of CO outgassing for the Centaurs (315898), (342842), and (382004), with $3\sigma$ upper limits all $Q({\rm CO}) < 10^{27}$ molecules s$^{-1}$.

However, there exists a small subset of solar system objects that are  enriched in CO. It is also not unprecedented for CO activity levels to change, or to be uncorrelated with dust production.  29P/Schwassmann-Wachmann 1 or SW1 is an active Centaur with a dynamically cold orbit with eccentricity, $e = 0.043$, and inclination, $i=9.37^\circ$, with $a=5.986$ au, a nucleus with radius $32\pm2$km \citep{Schambeau2018} and is on the verge of transitioning between the Centaur population and the JFCs \citep{Sarid2019}. 29P has had a dust coma for the past 90 years, and its outgassing is dominated by CO with $Q({\rm CO})$ between $1-7 \times 10^{28}$ molecules s$^{-1}$, more than a factor of 10 larger than any other measured species \citep{Senay1994,Crovisier1995,Gunnarsson2008,Paganini2013}. \citet{Wierzchos2020} presented observations of multiple CO and dust outbursts of 29P with the Arizona Radio Observatory 10 m Submillimeter Telescope during 2016 and 2018–2019, and found that the CO and dust outbursts were not  well correlated. They also reported that the CO production rate doubled during 2016 over the course of 70 hours, and reverted back within three days, which did not trigger a dust outburst. They concluded that  CO may not always be substantially incorporated with the dust component in the nucleus. CO was also detected in the active Centaur (60558) 174P/Echeclus at 6 au, with a rate of $Q({\rm CO}) =7.7\pm3.3 \times 10^{26}$ molecules s$^{-1}$ \citep{Wierzchos2017} and in (2060) Chiron $Q({\rm CO}) =2\pm1 \times 10^{28}$ molecules s$^{-1}$ \citep{Womack1997}. 174P/Echeclus is best known for exhibiting a massive outburst in 2005 when its brightness increased from V$\sim21$ to V$\sim14$ \citep{Choi2006,Bauer2008}, and subsequent smaller outbursts with magnitude changes of 3,3 and 4 in 2011, 2016, and 2017 \citep{Kareta2019,Jaeger2011,James2018}.

A particularly compelling object is the CO-dominated Comet C/2016 R2, an almost hyperbolic long period comet with an eccentricity of $e\sim0.997$ and $q\sim2.6AU$. \citet{Wierzchos2018} reported the first detection of CO in C/2016 R2 between between 2017 December and 2018 January with a production rate of $Q({\rm CO}) =4.6 \times 10^{28}$ molecules s$^{-1}$. \citet{McKay2019} presented observations when it was at around $2.8$au, and found that the production rate of CO was $Q({\rm CO}) = 9.5 \times 10^{28}$ molecules s$^{-1}$, and that the object was extremely dominated by CO with fractions of $Q({\rm CO}_2)/Q({\rm CO})\sim 18\%$ and $Q({\rm H_2O})/Q({\rm CO})\sim 0.3\%$. Most long period comets have much higher fractions of   H$_2$O and CO$_2$ to CO, and mostly greater than unity (for a reasonable comparison of the fractional compositions, see Table 7 in \citet{McKay2019}). The composition is so anomalous that \citet{McKay2019} speculated that R2 may even be of interstellar origin. While  production rates in comets show variability,  the production rate of CO has not been measured to vary by a factor of $\sim100$ in any individual object,  as required for the `Oumuamua scenario.


\section{Discussion}\label{sec:discussion}

In this paper, we investigated the spin dynamics of elongated small bodies with non-gravitational acceleration caused by either cometary outgassing or radiation pressure. We found that the spin state is steady for both mechanisms, so either one is consistent with observations obtained for `Oumuamua. Moreover, we applied our model to CO and found that it's energetically-allowable as `Oumuamua's source of propulsion. This composition poses difficulties however, given the non-detection of CO during the \textit{Spitzer} observations. If `Oumuamua was composed of CO, it is possible that activity started before it was detected.

If `Oumuamua's nucleus contains typical cometary substances, then its formation could be explained via conventional routes of minor body assembly. There exists an extensive literature of studies investigating the formation of planetesimals and planets in the solar system \citep{Chiang2010,Morbidelli2016}, and avenues such as the streaming instability \citep{Youdin2005} could form an object like `Oumuamua. 

Furthermore, if this hypothesis proves correct, then both interstellar objects discovered to date would be enriched in CO compared to most solar system small bodies. Because 2I/Borisov was larger than `Oumuamua and displayed a brilliant tail \citep{Jewitt2020}, it was detectable at much further heliocentric and geocentric distances. Borisov had a CO-rich tail \citep{Bodewits2020, Cordiner2020}, making this substance an \textit{a priori} compelling explanation of `Oumuamua's bulk composition. \citet{Kim2020} presented Hubble observations revealing that the changes in Borisov's coma morphology was evidence for seasonal effects, and that changes in activity levels and breakup \citep{Jewitt2020:BorisovBreakup} could be explained by regions on the northern hemisphere of the nucleus being exposed to the sun for the first time (particularly, see Figure 7 in \citet{Kim2020}). Moreover, Borisov's dust was consistent with grains mostly larger than $100\,\mu \text{m}$ \citep{Kim2020}. \citet{yang2021} presented VLT and ALMA observations that showed the presence of $\gtrsim1$mm size pebbles in the coma. They measured a CO production rate of Q$(CO)=3.3\pm0.8\times10^{26}\,\rm{s}^{-1}$ and found that the CO/H$_2$O mixing ratio changed drastically before and after perihelion, implying that the comet's home system experienced efficient radial mixing.  Moreover, comets in the Solar System undergo seasonal effects of outgassing. NASA's Deep Impact mission to 9P/Tempel 1 \citep{Ahearn2005} and  ESA's Rosetta Mission to the comet 67/Churyumov-Gerasimenko \citep{Glassmeier2007} demonstrated that the outgassing from sub-surface volatiles can cause frequent, and sometimes strong outbursts. Contextualized in conjunction with a smaller, sporadically-outgassing `Oumuamua, the two known interstellar objects would imply a diversity of shapes, sizes, and activity for interstellar comets.

While the specific composition of `Oumuamua is still undetermined, the forthcoming Vera Rubin Observatory's Legacy Survey of Space and Time (VRO/LSST) will place stricter constraints on the reservoir of interstellar objects. Given the population estimates from \cite{Do2018}, this campaign may see between $1-10$ interstellar objects per year \citep{Seligman2018, ivezic2019lsst}. Moreover, the ESA’s proposed \textit{Comet Interceptor} \citep{jones2019} mission or the NASA concept study \textit{BRIDGE} \citep{Moore2021} will be well-positioned to provide \textit{in situ} studies similar to the concept proposed in \citet{Seligman2018}.
 
\section{Acknowledgements}
We thank  Marco Micheli, Davide Farnocchia,  Ray Weymann, Julien de Wit, Adina Feinstein, Megan Mansfield, Christopher Lindsay, Emma Louden, and Konstantin Batygin for useful discussions. We thank the three anonymous reviewers for insightful comments and constructive suggestions that greatly strengthened the scientific content of this manuscript.

\bibliography{sample63}{}
\bibliographystyle{aasjournal}

\end{document}